\documentclass[epjCONF,columns]{svjour} %for 2 columns format
\usepackage{graphicx}
\usepackage[varg]{txfonts} % Times fonts
\usepackage[latin1]{inputenc}

\pdfoutput=1
\session-title{HCP 2011: Hadron Collider Physics Symposium 2011}
\begin{document}
\title{Determination of the top quark mass from the $t\bar t$  cross section \\ measured by CMS at $\sqrt{s} = 7$ TeV}
\author{M.~Aldaya\inst{1} \and K.~Lipka\inst{1} \and S.~Naumann-Emme\inst{1} for the CMS collaboration }
\institute{Deutsches Elektronen Synkrotron (DESY), Notkestrasse 85, 22607 Hamburg, Germany}
\abstract{
Higher-order QCD predictions are used to extract the top quark mass, both in the pole and in the $\overline{\mathrm{MS}}$ scheme, from the top quark pair production cross section measured in the dilepton final state. The analysed dataset corresponds to an integrated luminosity of 1.14~fb$^{-1}$ collected by the CMS experiment in 2011 in proton-proton collisions at $\sqrt{s} = 7$~TeV.
} %end of abstract
\maketitle
\section{Introduction}
\label{intro}
The top quark mass ($m_{t}$) is an important parameter of the Standard Model. Its precise measurement is one of the most relevant inputs to the global electroweak fits which provide constraints on the properties of the Higgs boson. 

Beyond leading-order (LO) Quantum Chromodynamics (QCD) predictions, $m_{t}$ depends on the renormalization scheme and its value can differ considerably for, e.g., pole mass or $\overline{\mathrm{MS}}$ mass definitions. It is therefore important to understand how to interpret the experimental result in terms of the renormalization conventions.

Direct measurements of the top quark mass from the reconstruction of the final states in top-antitop ($t \bar t$) decays rely highly on the detailed description of the corresponding signal in Monte Carlo (MC) simulations. The quantity measured in the data is compared to the simulation and thus corresponds to the top quark mass definition used in the MC generator ($m_{t}^\mathrm{MC}$). All currently used MC simulations contain only matrix elements at fixed order (leading or next-to-leading order) QCD, while higher orders are simulated by applying parton showers. Therefore, they are not precise enough to fix the renormalization scheme, which leads to an uncertainty in the input top quark mass definition.

We extract the top quark mass by comparing the inclusive $t \bar t$ production cross section, $\sigma_{t\bar t}$, in the dilepton channel measured by CMS~\cite{TOP-11-005} to fully-inclusive calculations at higher-order QCD that involve an unambiguous definition of $m_{t}$. This extraction provides an important test of the mass scheme as applied in MC simulations and gives complementary information, with different sensitivity to theoretical and experimental uncertainties than the direct measurements of $m_{t}^\mathrm{MC}$ which rely on the kinematic details of the mass reconstruction.

\section{Method}
\label{sec:1}

The combined measured ${t\bar t}$ cross section in dilepton decays~\cite{TOP-11-005} corresponding to an integrated luminosity of 1.14~fb$^{-1}$ of collected data, $\sigma_{t\bar{t}}=169.90 \pm 3.9$ (stat.) $\pm 16.3$ (syst.) $\pm 7.6$ (lumi.)$\;\mathrm{pb}$, is used to determine the mass of the top quark through its comparison to different higher-order QCD predictions using the pole and the $\overline{\mathrm{MS}}$ mass definitions. For the measured cross section it is assumed that $m_{t}^\mathrm{MC} \equiv m_{t}^\mathrm{pole}$. The dependence of $\sigma_{t\bar{t}}$ on the assumed value of $m_{t}^\mathrm{MC}$ via the signal acceptance is studied using different simulated samples of ${t\bar t}$ events generated at different values of $m_{t}^\mathrm{MC}$ and parametrized in a polynomial form for each of the dilepton decay channels. The weighted average of the parametrizations is used to obtain the central value of $m_t$~\footnote{Here and in the following, $m_t$ denotes the top quark mass in both the pole and the $\overline{\mathrm{MS}}$ definition.}. 

%---------------------------------
\begin{figure}[h]
  \centering
  \includegraphics[width=0.47\textwidth]{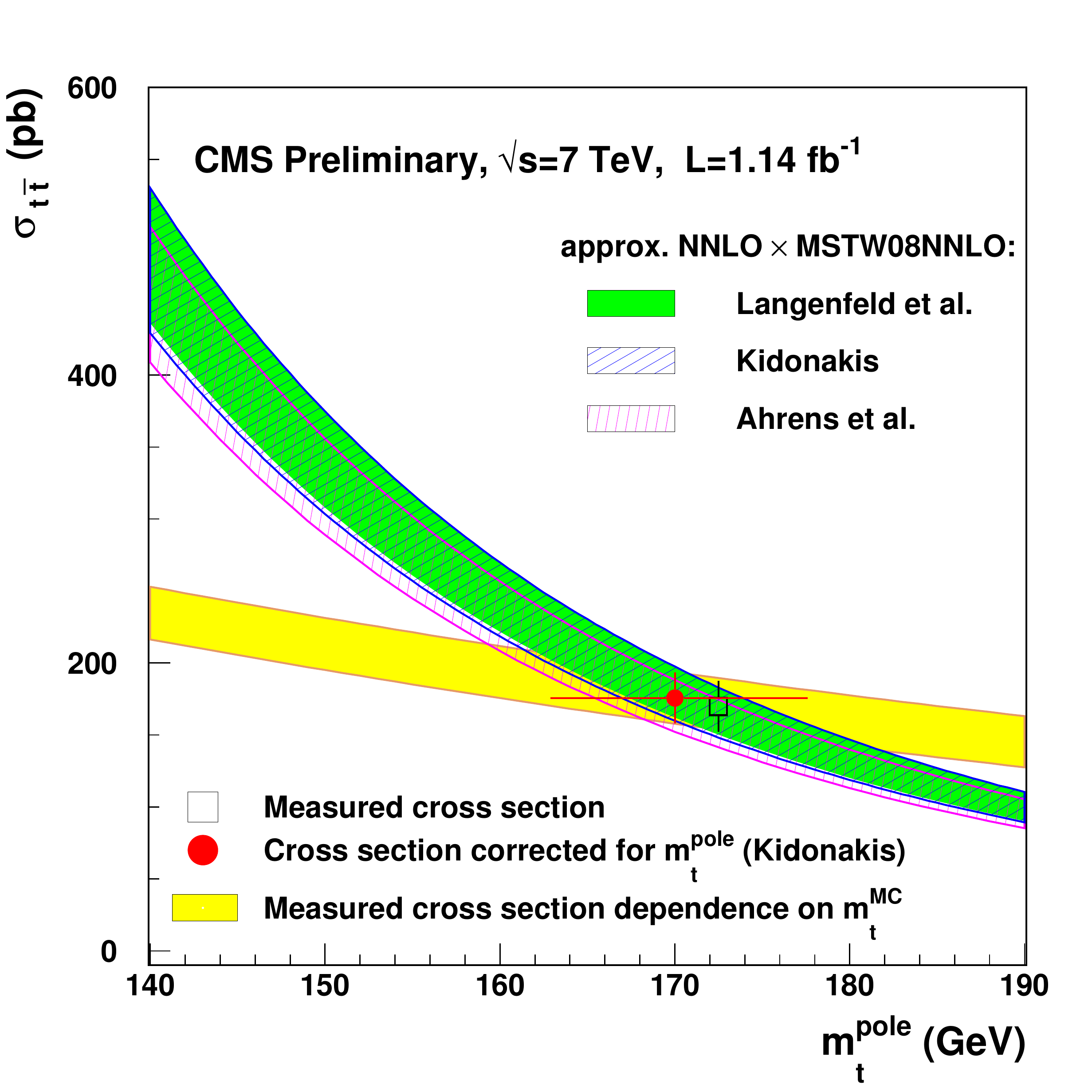} 
  \caption{Comparison of the predicted cross section $\sigma_{t\bar t}(m_t^\mathrm{pole})$ for the different calculations and the measured cross section (open square) as a function of $m_t^\mathrm{pole}$. The measured $\sigma_{t\bar{t}}$ corrected for the $m_t^\mathrm{pole}$ dependence at the determined value of $m_t^\mathrm{pole}$ using the calculation~\cite{Kidonakis} is also depicted (filled circle). The dependence of the measured $\sigma_{t\bar{t}}$ on $m_{t}^\mathrm{MC}$ is shown as a light shaded band. Different approximate NNLO predictions are given as differently hatched bands.} 
  \label{fig:figure01}
\end{figure}

%----------------------------------

Three different approaches~\cite{Moch,Kidonakis,Ahrens} to calculate the higher-order corrections to the next-to-leading order (NLO) calculations of ${t\bar t}$ production (approximate NNLO) have been used to extract $m_t$. The top quark mass is given in both the pole and $\overline{\mathrm{MS}}$ definitions in the calculations~\cite{Moch,Ahrens}, while it is defined as a pole mass in the prediction~\cite{Kidonakis}. The calculations are performed using the MSTW08NNLO~\cite{mstw08nnlo} parton distribution functions (PDF). 

The uncertainty of the approximate NNLO calculations includes the error due to variation of renormalization and factorisation scales ($\mu_r$ and $\mu_f$, respectively), the uncertainty on the parton luminosity added in quadrature and the uncertainty due to variation of the strong coupling constant ($\alpha_S(M_Z)$) in the PDF~\cite{mstw08nnlo}. 

The top quark mass and its uncertainty are determined using a joint-likelihood approach, similar to~\cite{D0mass,ATLAS-CONF-2011-054}. The probability density function $f_{exp} (\sigma_{t\bar{t}} | m_t)$ is constructed from a Gaussian distribution with the measured cross section as its mean value, using a third-order polynomial to parametrize the mass dependence, and the total experimental uncertainty as the width of this Gaussian. The probability density function $f_{th} (\sigma_{t\bar{t}} | m_t)$ is constructed from a Gaussian centered on the predicted cross section, with the standard deviation adjusted to the uncertainty of the prediction. For this purpose, the predictions were parametrized as 
{
\setlength\abovedisplayshortskip{-5pt}
\setlength\belowdisplayshortskip{-1pt}
$$\sigma_{t\bar{t}} (m_t) = \frac{1}{m_t^4} \left( a + b \cdot m_t + c \cdot m_t^2 + d \cdot m_t^3  \right).$$
}

Finally, $m_t$ is obtained as the maximum of a combined uncorrelated likelihood constructed as 
{
\setlength\abovedisplayshortskip{-3pt}
\setlength\belowdisplayshortskip{-1pt}
$$L(m_t) = \int f_{exp} (\sigma_{t\bar{t}} | m_t) f_{th} (\sigma_{t\bar{t}} | m_t) d\sigma_{t\bar{t}}.$$
}

Asymmetric uncertainties on $m_t$ are determined from the 68.3\% area around 
the maximum that yields equal probabilities at its left and right edges. In order to estimate the uncertainty arising from the different parametrizations for the measured $\sigma_{t\bar{t}}$ determined for each dilepton channel, the analysis is repeated for all three parametrizations and the spread of the resulting mass values is taken as an uncertainty, which is added in quadrature to the uncertainty obtained from the shape of the original joint likelihood. To evaluate the uncertainty by defining $m_t^\mathrm{MC}=m_t^\mathrm{pole}$, a shift of $m_t^\mathrm{MC}$ by $\pm$1~GeV is applied and the resulting shift of the extracted mass is also added quadratically. 

\section{Results and conclusions}

In Fig.~\ref{fig:figure01}, the measured cross section together with its dependence on the $m_t^\mathrm{MC}$ assumption and approximate NNLO predictions are presented. As an example, the value of the measured cross section corrected for the extracted $m_t^\mathrm{pole}$ using the calculation~\cite{Kidonakis} is also depicted. 

\begin{figure}[ht]
\centering
  \includegraphics[width=0.45\textwidth]{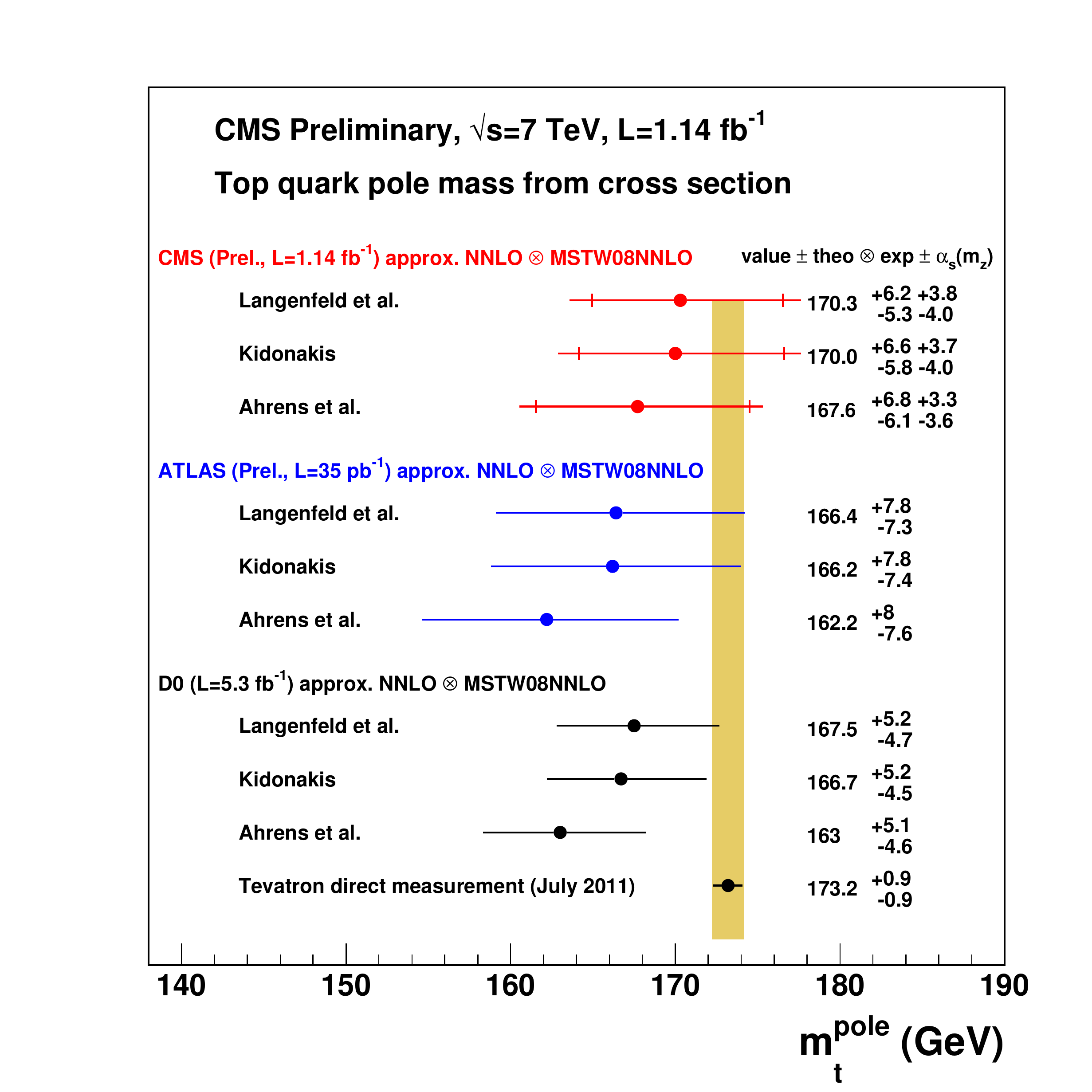}
  \includegraphics[width=0.45\textwidth]{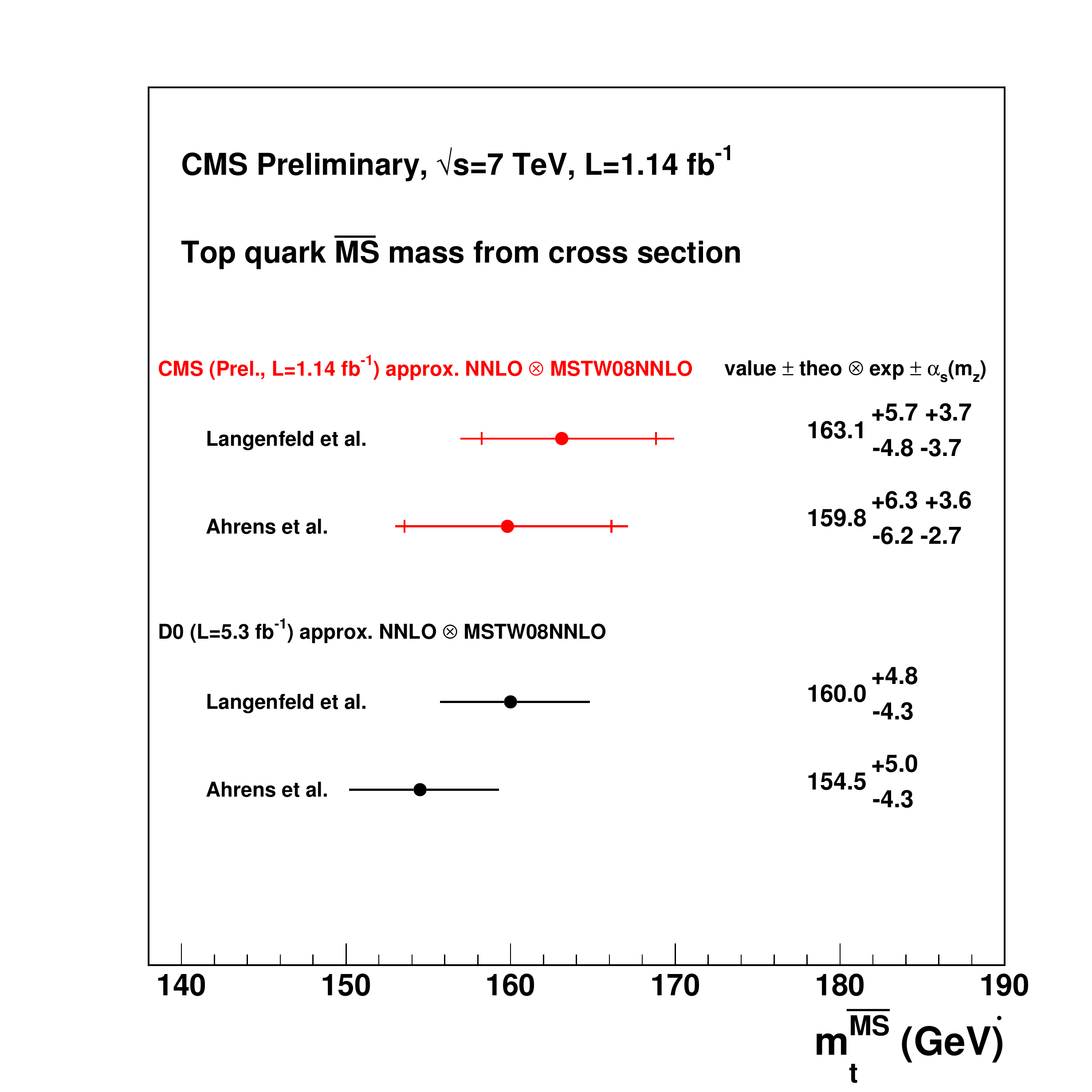}
  \caption{Top quark pole (upper) and $\overline{\mathrm{MS}}$ mass (lower) extracted from the measured $\sigma_{t\bar{t}}$ compared to results from~\cite{D0mass,ATLAS-CONF-2011-054} and the direct top quark mass world average~\cite{TevatronComb}. 
The inner error bar on the CMS result includes the experimental, PDF and scale variation uncertainty, the outer one corresponds to the uncertainty including variation of $\alpha_S(M_Z)$ in the PDF.}  
  \label{fig:figure02}
\end{figure}

The values of the top pole mass obtained using the three approximate NNLO predictions are summarized in Table~\ref{tab:mstw}. In Fig. \ref{fig:figure02} (upper), the same values are compared to similar measurements by the ATLAS~\cite{ATLAS-CONF-2011-054} and D0~\cite{D0mass} collaborations. These results are in very good agreement. They are consistently lower than the average of the direct measurements at Tevatron~\cite{TevatronComb}, which is also shown.

The theoretical calculations~\cite{Moch,Ahrens} are also available as a function of $m_t^{\overline{\mathrm{MS}}}$. For the measured cross section, it is again assumed that $m_{t}^\mathrm{MC}$ corresponds to $m_{t}^\mathrm{pole}$ and this is then translated into $m_t^{\overline{\mathrm{MS}}}$ using the relation at the three-loop level~\cite{Melnikov2000qh,Bethke2009jm}. The extracted values of $m_t^{\overline{\mathrm{MS}}}$ are also listed in Table~\ref{tab:mstw} and compared in Fig.~\ref{fig:figure02}~(lower) to those obtained by D0~\cite{D0mass}.

Both the experimental and the theoretical uncertainty contribute similarly to the uncertainty of the top quark mass determination, with the dominant uncertainty on the approximate NNLO calculations being the variation of the $\alpha_S(M_Z)$ in the PDF, so far not accounted for by the previous analyses~\cite{D0mass,ATLAS-CONF-2011-054}.

Alternatively, the top quark mass was obtained using the calculations~\cite{Moch} and~\cite{Ahrens} with HERAPDF15NNLO~\cite{hera15nnlo,herapdf_alfas}, taking into account the experimental PDF uncertainty, variation of $\mu_r$ and $\mu_f$ and of the $\alpha_S(M_Z)$ value. The resulting values for $m_t^\mathrm{pole}$ and $m_t^{\overline{\mathrm{MS}}}$ are listed in Table~\ref{tab:hera}. They are higher by 1.2-1.5~GeV compared to the results obtained with MSTW08NNLO (c.f Table~\ref{tab:mstw}) due to the different values of $\alpha_S(M_Z)$ that are used in the respective PDF sets. The uncertainties on the extracted masses are slightly smaller when using HERAPDF15NNLO. In addition, the HERAPDF approach allows for various different studies of the influence of PDF fit assumptions on $\sigma_{t\bar{t}}$.  

\begin{table}[htbp]
    \caption{\small 
      Top quark mass in the pole and $\overline{\mathrm{MS}}$ definition as obtained using the measured $\sigma_{t\bar{t}}$ and approximate NNLO calculations with MSTW08NNLO PDF.}  
    \label{tab:mstw} 
    \begin{tabular}{lll}
     \hline\noalign{\smallskip}
      Approx. NNLO $\times$ MSTW08NNLO & $m_t^{pole}$ / GeV  & $m_t^{\overline{MS}}$ / GeV\\ 
     \noalign{\smallskip}\hline\noalign{\smallskip}
      Langenfeld et al.~\cite{Moch}    & $170.3^{+7.3}_{-6.7}$ & $163.1^{+6.8}_{-6.1}$\\ 
      Kidonakis~\cite{Kidonakis}       & $170.0^{+7.6}_{-7.1}$ & -- \\ 
      Ahrens et al.~\cite{Ahrens}      & $167.6^{+7.6}_{-7.1}$ & $159.8^{+7.3}_{-6.8}$\\
      \noalign{\smallskip}\hline 
    \end{tabular}
\end{table}

\begin{table}[htbp]
    \caption{\small 
      Top quark mass in the pole and $\overline{\mathrm{MS}}$ definition as obtained using the measured $\sigma_{t\bar{t}}$ and approximate NNLO calculations with HERAPDF15NNLO PDF.}  
    \label{tab:hera} 
    \begin{tabular}{lll}
     \hline\noalign{\smallskip}
      Approx. NNLO $\times$ HERAPDF15NNLO & $m_t^{pole}$ / GeV  & $m_t^{\overline{MS}}$ / GeV\\ 
     \noalign{\smallskip}\hline\noalign{\smallskip}
      Langenfeld et al.~\cite{Moch}    & $171.7^{+6.8}_{-6.0}$ & $164.3^{+6.5}_{-5.7}$\\ 
      Ahrens et al.~\cite{Ahrens}      & $169.1^{+6.7}_{-5.9}$ & $161.0^{+6.8}_{-6.1}$ \\ 
      \noalign{\smallskip}\hline 
    \end{tabular}
\end{table}

To summarize, we extract the top quark pole and $\overline{\mathrm{MS}}$ mass by comparing the measured $\sigma_{t\bar{t}}$ with different higher-order QCD calculations~\cite{TOP-11-008}. The measurement was performed by the CMS collaboration in $t\bar t$ decays with dileptonic final states using 1.14~fb$^{-1}$ of collected data at $\sqrt{s}=7$~TeV. The results are in very good agreement with similar measurements by D0 and ATLAS and provide the first determination of the top quark pole and $\overline{\mathrm{MS}}$ mass at CMS. The uncertainty of the results is dominated by the systematic error of the measured $\sigma_{t\bar{t}}$ and by the PDF uncertainty in the theory, including the variation of $\alpha_S(M_Z)$ in the PDF.

\end{document}